\documentclass{elsart}

\usepackage{HEP} 
\usepackage{amsmath}
\usepackage{amssymb}
\usepackage{amsfonts}
\usepackage{amscd}
\usepackage{bbm} 
\usepackage{fleqn} 
\usepackage[footnotesize]{caption}
\usepackage{graphicx} 
\usepackage{braket} 
\usepackage{mathrsfs}
\usepackage[center,footnotesize,hang]{subfigure}
\usepackage[bbgreekl]{MyBbol}
\usepackage{longtable}

\usepackage{accents}

\newcommand{\CenterObject}[1]{\vcenter{\hbox{#1}}}

\def\D{\mathrm{d}} 
\def\SU{\text{SU}}
\def\SO{\text{SO}}
\def\U{\text{U}}

\def\<{\left\langle}
\def\>{\right\rangle}

\def\Ng{j}	
\def\ChargeC{\mathrm{C}}

\newcommand{\SuperField}[1]{\bbsymbol{#1}}  
\newcommand{\RaiseBrace}[1]{\raise3pt\hbox{$\displaystyle#1$}}

\DeclareMathOperator{\Tr}{Tr}
\DeclareMathOperator{\diag}{diag}

\allowdisplaybreaks[1]

\begin{document}

\begin{frontmatter}

\begin{flushright}

{\small TUM-HEP-459/02}

\end{flushright}
\vspace*{2.cm}
\title{
 Neutrino Mass Matrix Running for Non-Degenerate See-Saw Scales
}
\author{Stefan Antusch\thanksref{label01}},
\thanks[label01]{E-mail: \texttt{santusch@ph.tum.de}}
\author{J\"{o}rn Kersten\thanksref{label04}},
\thanks[label04]{E-mail: \texttt{jkersten@ph.tum.de}}
\author{Manfred Lindner\thanksref{label05}}
\thanks[label05]{E-mail: \texttt{lindner@ph.tum.de}}
and 
\author{Michael Ratz\thanksref{label06}
}
\thanks[label06]{E-mail: \texttt{mratz@ph.tum.de}}
\address{Physik-Department T30, 
Technische Universit\"{a}t M\"{u}nchen\\ 
James-Franck-Stra{\ss}e,
85748 Garching, Germany
}

\begin{abstract}
We consider the running of the neutrino mass matrix
in the Standard Model and the Minimal Supersymmetric Standard Model,
extended by heavy singlet Majorana neutrinos.
Unlike previous studies, we do not assume that all of the heavy
mass eigenvalues are degenerate.  This leads to various effective
theories when the heavy degrees of freedom are integrated out
successively.
We calculate the Renormalization Group Equations that govern 
the evolution of the neutrino mass matrix in these effective theories.
We show that an appropriate treatment of the singlet mass scales can 
yield a substantially different result compared to integrating out the
singlets at a common intermediate scale.
\end{abstract}

\end{frontmatter}

\newpage

\section{Introduction}
The discovery of neutrino masses requires an extension of the Standard Model
(SM) or the Minimal Supersymmetric Standard Model (MSSM), which may involve 
right-handed neutrinos, or more generally gauge singlets. 
Since there are no protective symmetries, 
these singlets are usually expected to have huge explicit (Majorana) masses.
This leads to the see-saw mechanism \cite{seesaw}, which provides a convincing explanation 
for small neutrino masses.
This scenario can be realized in many Grand Unified Theories (GUT's) and
their supersymmetric counterparts.  For instance,
left-right symmetric models and $\SO (10)$ GUT's
include singlet neutrinos, which can get huge masses
in several ways, e.g.\ by a Higgs in a suitable representation
or radiatively.
Furthermore, additional singlets may exist,
which can also be involved in the see-saw mechanism. 

It is often assumed that all heavy singlet mass eigenvalues are degenerate.
However, in all the models a large hierarchy 
of the singlet masses is possible.
Note that such a hierarchical spectrum may even show up if all elements
of the singlet mass matrix are of the same order.  Democratic mass
matrices, where this is the case due to 
discrete symmetries, are an example.  
Another argument for a non-degenerate spectrum follows from assuming
a neutrino Yukawa matrix $Y_\nu$ which is proportional to the
diagonalized charged lepton Yukawa matrix $Y_e$, i.e.\ the relation 
$Y_\nu = c_\nu Y_e \approx c_\nu \diag(10^{-2},10^{-3},10^{-5})$ holds 
with a constant real number $c_\nu$.
If the neutrino masses are degenerate and of the order $1\,\mathrm{eV}$,
the see-saw relation 
$\kappa = \frac{4}{v^2_\mathrm{EW}} M_\nu = 2 Y_\nu^T M^{-1} Y_\nu$
for the neutrino mass matrix $M_\nu$ allows 
to determine the singlet mass matrix $M$.  Mixings do not significantly
alter this picture, since e.g.\ bimaximal mixing can be accomplished by
small modifications of a degenerate $M_\nu$ of the order
$10^{-2}\,\mathrm{eV}$ or $10^{-3}\,\mathrm{eV}$, respectively. 
Taking for example $c_\nu = 100$, 
the mass eigenvalues of $M$ are of the order
$10^{7}\,\mathrm{GeV}$, $10^{11}\,\mathrm{GeV}$ and 
$10^{13}\,\mathrm{GeV}$ for the case at hand.
It is therefore conceivable that there may
be an even larger hierarchy in $M$ than in
the charged lepton Yukawa matrices. 
Altogether, there are thus good reasons to study the effects of a
non-degenerate or even hierarchical singlet mass spectrum.

In this paper, we calculate the Renormalization Group Equations 
(RGE's) for the evolution of the neutrino mass matrix from the GUT scale to the
electroweak or SUSY breaking scale. We consider the case 
where the SM and the MSSM are extended by an arbitrary number of 
heavy singlets which have explicit (Majorana) masses with
a non-degenerate spectrum. 
Hence, to study the RG evolution of neutrino masses several 
Effective Field Theories (EFT's), with the singlets partly integrated
out, have to be taken into account. Below the lowest mass threshold, the
neutrino mass matrix is given
by the effective dimension 5 neutrino mass operator in the SM or MSSM, 
respectively.  The corresponding RGE's were derived in 
\cite{Chankowski:1993tx,Babu:1993qv,Antusch:2001ck,Antusch:2002vn,Antusch:2002ek}.

\section{Effective Theories from Integrating Out Singlet Neutrinos}
\label{sec:SMandMSSMAddSingletMaj}
\begin{figure}[h]
 \begin{center}
  \vspace{5mm}
  \includegraphics{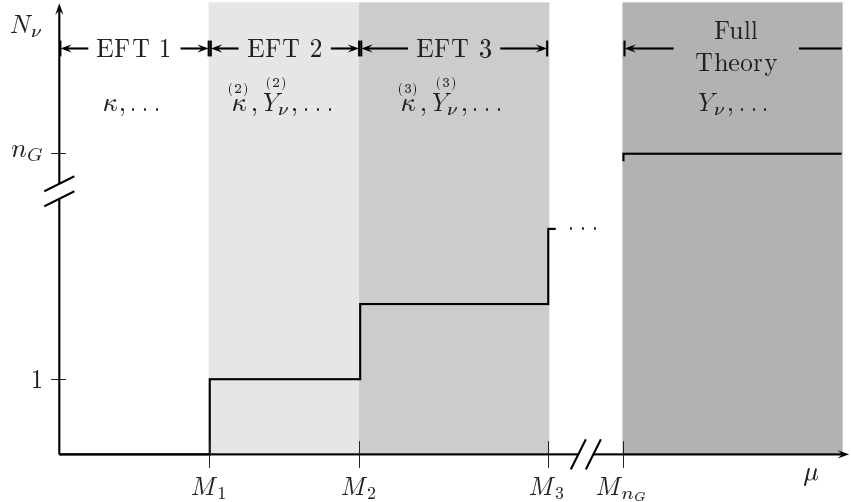}
 \end{center}
 \caption{Illustration of the ranges of the different theories.
 The EFT's emerge from successively integrating out the heavy fields.
 ``EFT 1'' corresponds to the SM or MSSM with additional dimension 5
 mass operators for neutrino masses. 
 ``Full Theory'' refers to the SM or MSSM, extended by $n_G$ gauge singlets.
 The meaning of the variables is explained in the text. 
}
 \label{fig:EFTRanges}
\end{figure}

Consider the SM or the MSSM with $n_G$ additional sterile neutrinos. 
The eigenvalues of the mass matrix \(M\), i.e.\ the masses of the mass
eigenstates $\{N_\mathrm{R}^1,\dots,N_\mathrm{R}^{n_G}\}$,
have a certain spectrum, $M_1 \le M_2 \le \dots \le M_{n_G}$.
We will consider the general case that this spectrum is non-degenerate.
Successively integrating out the heavy sterile neutrinos at the
thresholds $M_i$ results in effective theories, valid in certain
energy ranges as depicted in figure~\ref{fig:EFTRanges}.

Before we calculate the RGE's in the various theories, let us specify
the modifications in the Lagrangians due to the appearance of the heavy
neutrinos.
In the SM above the highest mass threshold (``Full Theory'' in figure
\ref{fig:EFTRanges}), the kinetic and mass term as well as the Yukawa
interaction for the singlet neutrinos 
$N_\mathrm{R}^i$, $i\in\{1,\dots,n_G\}$, 
are added: 
\begin{eqnarray}
\mathscr{L}_\mathrm{N} = 
 \overline{N_\mathrm{R}^i} (i \gamma^\mu \partial_\mu) N_\mathrm{R}^i
+\left( -\frac{1}{2} \overline{N_\mathrm{R}^i} M_{ij} N^{\ChargeC j}_\mathrm{R}
-(Y_{\nu})_{if}\overline{N_\mathrm{R}^i} \widetilde \phi^\dagger \ell^f_\mathrm{L}
+\text{h.c.} \right) \;,
\end{eqnarray}
where $N_\mathrm{R}^\ChargeC := (N_\mathrm{R})^\ChargeC$ is the
charge conjugate of $N_\mathrm{R}$.  $f\in\{1,\dots,n_F\}$ are flavour
indices, $\ell_\mathrm{L}^f$ are the SU(2)$_\mathrm{L}$-doublets of
leptons, $\phi$ is the Higgs doublet, and 
$\widetilde \phi := i \sigma^2 \phi^*$.
Summation over repeated indices is implied throughout the paper.
For the calculation of the RGE's, we will work in a basis 
in which the Majorana mass matrix \(M\) is diagonal. 

In the MSSM, the additional gauge singlet 
Weyl spinors $\nu^{\mathrm{C}i}$, which correspond to the
right-handed Dirac spinors $N_\mathrm{R}^i$, and their superpartners
are components of
the chiral superfields $\SuperField{\nu}^{\mathrm{C}i}$.
The terms of the superpotential containing these superfields are     
\begin{eqnarray}
\mathscr{W}_\mathrm{(N)} = \frac{1}{2} \SuperField{\nu}^{\mathrm{C}i} M^{}_{ij}
\SuperField{\nu}^{\mathrm{C}j}
+(Y^{}_{\nu})_{if} \SuperField{\nu}^{\mathrm{C}i} \SuperField{h}_a^{(2)}
(\varepsilon^T)^{ab} \SuperField{l}_b^f + \text{h.c.} \;,
\end{eqnarray}
where $\SuperField{l}^{f}$ and $\SuperField{h}^{(2)}$ are the chiral
superfields that contain the leptonic SU(2)$_\mathrm{L}$-doublets and
the Higgs doublet with weak hypercharge $+\frac{1}{2}$.
$\varepsilon$ is the totally antisymmetric tensor in 
2 dimensions, and $a,b,c,d \in \{1,2\}$ are SU(2) indices.

The Higgs doublet superfield $\SuperField{h}^{(1)}$ with weak
hypercharge $-\frac{1}{2}$ is involved
in the Yukawa couplings of the \(\mathrm{SU}(2)_\mathrm{L}\)-singlet 
superfields $\SuperField{e}^{\mathrm{C}}$ and 
$\SuperField{d}^{\mathrm{C}}$ containing the charged
leptons and down-type quarks, whereas 
$\SuperField{h}^{(2)}$ 
couples to $\SuperField{\nu}^{\mathrm{C}}$ and the 
superfield $\SuperField{u}^{\mathrm{C}}$ containing the up-type quarks.
The part of the superpotential describing the remaining
Yukawa interactions is given by
\begin{eqnarray}
 \mathscr{W}_{\mathrm{Yuk}}^{\mathrm{MSSM}} 
 &=&
 (Y_e)_{gf}\SuperField{e}^{\ChargeC g}
 	\SuperField{h}^{(1)}_a\varepsilon^{ab}\SuperField{l}^f_b
 \nonumber \\
 && {}
 +(Y_d)_{gf}\SuperField{d}^{\ChargeC g}
 	\SuperField{h}^{(1)}_a\varepsilon^{ab}\SuperField{q}^f_b
 +(Y_u)_{gf}\SuperField{u}^{\ChargeC g}
 \SuperField{h}^{(2)}_a (\varepsilon^T)^{ab}  \SuperField{q}_b^f \;,
\end{eqnarray}
where $\SuperField{q}$ is the quark doublet superfield.
The field content of the superfields is
\begin{subequations}\label{eq:superfieldsMSSM}
\begin{eqnarray}
\SuperField{l}^f &=& \widetilde{\ell} ^f + 
\sqrt{2}\, \theta \ell^f + \theta\theta\,\,F_\ell^f\;, \\
\SuperField{e}^{\ChargeC g} &=& \widetilde{e}^{\ChargeC g} 
+ \sqrt{2}\, \theta e^{\ChargeC g} + \theta\theta\,\,F_e^g\;, \\ 
 \SuperField{\nu}^{\ChargeC \Ng} &=& \widetilde{\nu}^{\ChargeC \Ng} 
+ \sqrt{2}\, \theta \nu^{\ChargeC \Ng} + \theta\theta\,\,F_\nu^\Ng \; , \\
\SuperField{q}^f &=& \widetilde{q} ^f + \sqrt{2}\, \theta q^f
  + \theta\theta\,\,F_q^f\;, \\
 \SuperField{u}^{\ChargeC g} &=& \widetilde{u}^{\ChargeC g} 
+ \sqrt{2}\, \theta u^{\ChargeC g} + \theta\theta\,\,F_u^g \;,\\
 \SuperField{d}^{\ChargeC g} &=& \widetilde{d}^{\ChargeC g} 
+ \sqrt{2}\, \theta d^{\ChargeC g} + \theta\theta\,\,F_d^g \;,\\
\SuperField{h}^{(1)} &=& \phi^{(1)} + \sqrt{2}\, \theta 
\Tilde{\phi}^{(1)} + 
\theta\theta\,\,F_{h^{(1)}} \;,\\
\SuperField{h}^{(2)} &=& \phi^{(2)} + \sqrt{2}\, \theta 
\Tilde{\phi}^{(2)} + 
\theta\theta\,\,F_{h^{(2)}} \;.
\end{eqnarray}
\end{subequations}

By integrating out all singlet neutrinos of the extended SM, one obtains
the dimension 5 operator that gives Majorana masses to the light neutrinos,
\begin{equation}\label{eq:Kappa}
 \mathscr{L}_{\kappa}^{\mathrm{SM}}  
 =\frac{1}{4} 
 \kappa_{gf} \, \overline{\ell_\mathrm{L}^\mathrm{C}}^g_c\varepsilon^{cd} \phi_d\, 
 \, \ell_{\mathrm{L}b}^{f}\varepsilon^{ba}\phi_a  
  +\text{h.c.} \;.
\end{equation}

The corresponding expression in the MSSM is the $F$-term of
\begin{equation}\label{eq:Kappa-MSSM}
 \mathscr{W}_{\kappa}^{\mathrm{MSSM}} 
 =-\frac{1}{4} 
  {\kappa}^{}_{gf} \, \SuperField{l}^{g}_c\varepsilon^{cd}
 \SuperField{h}^{(2)}_d\, 
 \, \SuperField{l}_{b}^{f}\varepsilon^{ba} \SuperField{h}^{(2)}_a 
 +\text{h.c.} \;.
\end{equation}

In the intermediate region between the $(n\!-\!1)$th and the $n$th
threshold, the singlets $\{N_\mathrm{R}^n,\dots,N_\mathrm{R}^{n_G}\}$ or
singlet superfields
$\{\SuperField{\nu}^{\ChargeC n},\dots,\SuperField{\nu}^{\ChargeC n_G}\}$
are integrated out, leading to an effective operator of the type
\eqref{eq:Kappa} or \eqref{eq:Kappa-MSSM} with coupling constant
$\accentset{(n)}{\kappa}_{gf}$, where $\accentset{(1)}{\kappa}_{gf}$ is
identical to $\kappa_{gf}$.
In this region, the Yukawa matrix for the remaining singlet
neutrinos is a $(n\!-\!1) \times n_F$ matrix and will be referred to as
$\accentset{(n)}{Y}_\nu$, 
\begin{equation}
 Y_\nu\to
 \left(
  \begin{array}{ccc}
   	(Y_\nu)_{1,1} & \cdots & (Y_\nu)_{1,n_F} \\
    \vdots & & \vdots \\
 	(Y_\nu)_{n-1, 1} & \cdots & (Y_\nu)_{n-1, n_F}\vspace*{1mm} \\
	\noalign{\hrule\vspace*{1mm}}
		0 & \cdots & 0\cr
	\vdots & & \vdots \vspace*{1mm}\cr
	0 & \cdots & 0 \cr
  \end{array}
 \right)
 \:
 \begin{array}{cl}
  \left.\begin{array}{c}\\[1.7cm]\end{array}\right\}
  & =: \accentset{(n)}{Y}_\nu\;,
  \\
  \left.\begin{array}{c}\\[1.7cm]\end{array}\right\}
  	&  \begin{array}{l}n_G\!-\!n\!+\!1\:\text{heavy, sterile}\\[-0.3cm]
  \text{neutrinos integrated out}\;.\end{array}
 \end{array}
\end{equation}

The tree-level matching condition for the effective coupling constant
at the threshold corresponding to the largest eigenvalue $M_n$ of
$\accentset{(n+1)}{M}$ is given by
\begin{equation} \label{eq:MatchingConditionKappaN}
	\accentset{(n)}{\kappa}_{gf} \big|_{M_n} :=
	\hphantom{^{(n}} \accentset{(n+1)}{\kappa}_{gf} \big|_{M_n} +
	2\,\RaiseBrace{\bigl(} \;\; {\accentset{(n+1)}{Y}_\nu}^{\;T}
	\RaiseBrace{\bigr)}_{gn} M_n^{-1}
	 \RaiseBrace{\bigl(} \;\; \accentset{(n+1)}{Y}_\nu \;\, 
	 \RaiseBrace{\bigr)}_{nf} \big|_{M_n} 
	\;\; \text{(no sum over $n$).}
\end{equation}
To determine the RGE's, we first calculate the relevant counterterms for
the effective theories.  We use dimensional regularization (with
$d := 4\!-\!\epsilon$ dimensions) and the MS renormalization scheme.
The renormalization constants below the $n$th threshold are denoted by 
$\accentset{(n)}{Z}$, $\delta\accentset{(n)}{\kappa}$, etc., analogous
to our notation for the coupling constants.

\section{Calculation of the Counterterms}

For the one-loop wavefunction renormalization constants 
$\accentset{(n)}{Z} := \mathbbm{1} + \delta\accentset{(n)}{Z}$ between
the thresholds in the extended SM, we find in $R_\xi$
gauge for $\U(1)_\mathrm{Y}$ and $\SU(2)_\mathrm{L}$
\begin{subequations}
\begin{eqnarray}
\delta\accentset{(n)}{Z}_{\ell_\mathrm{L}}
& = & -\frac{1}{16\pi^2}  \left[ 
 \accentset{(n)}{Y}^\dagger_\nu 
\accentset{(n)}{Y}_\nu 
+ Y_e^\dagger Y_e + \frac{1}{2}\xi_B g^2_1 + \frac{3}{2}\xi_W g^2_2\right]\frac{1}{\epsilon}\;,
\\
\delta\accentset{(n)}{Z}_{\phi}
& = & 
-\frac{1}{16\pi^2}\left[
2\,\Tr \RaiseBrace{\bigl(}\accentset{(n)}{Y}^{\dagger}_\nu \accentset{(n)}{Y}_\nu\RaiseBrace{\bigr)} 
+2\,\Tr (Y_e^\dagger Y_e) 
+6\,\Tr (Y_u^\dagger Y_u)  
+6\,\Tr (Y_d^\dagger Y_d)  
\right. \nonumber \\
&&\left. \hphantom{-\frac{1}{16\pi^2}\left[\right.}+ 
	\frac{1}{2}(\xi_B - 3) g^2_1 + 
	\frac{3}{2}(\xi_W - 3) g^2_2 
\right]\frac{1}{\epsilon} \;,
\\
\delta\accentset{(n)}{Z}_{N}
& = & 
-\frac{1}{16\pi^2}\left[
2\, \accentset{(n)}{Y}_\nu 
\accentset{(n)}{Y}^\dagger_\nu 
\right]\frac{1}{\epsilon} \;.
\end{eqnarray}
\end{subequations}
For the vertex renormalization constants we obtain
\begin{subequations}
\begin{eqnarray}
\delta\accentset{(n)}{Y_\nu} 
&=& -  \frac{1}{16\pi^2} \left[
2\, \accentset{(n)}{Y_\nu}\, (Y^\dagger_e Y_e) + 
\frac{1}{2} \xi_B g^2_1\,\accentset{(n)}{Y_\nu}
+\frac{3}{2} \xi_W g^2_2 \,\accentset{(n)}{Y_\nu}\right]\frac{1}{\epsilon}\;, 
\\
\delta\accentset{(n)}{\kappa} 
&=&
-\frac{1}{16\pi^2}
 \left[2\,(Y_e^\dagger Y_e)^T\:\accentset{(n)}{\kappa} 
 + 2\,\accentset{(n)}{\kappa}\,(Y_e^\dagger Y_e)
 	-\lambda \accentset{(n)}\kappa
	\vphantom{\frac{1}{2}} \right.
\nonumber \\
&& \left.\hphantom{-\frac{1}{16\pi^2} \left[\right.} 
	+\frac{1}{2}(2\xi_B - 3) g_1^2 \: \accentset{(n)}{\kappa} + 
	\frac{3}{2}(2\xi_W-1) g_2^2 \: \accentset{(n)}{\kappa} \,\right]
	\frac{1}{\epsilon} \;,
\\
\delta\accentset{(n)}{M} &=& 0 \;,
\end{eqnarray} 
\end{subequations}
where $\lambda$ is the scalar quartic coupling appearing in the
interaction term $-\frac{1}{4} \lambda (\phi^\dagger \phi)^2$.
The above quantities are defined by the counterterms for the mass and
the Yukawa vertex of the sterile neutrinos as well as the one for the
effective vertex,
\begin{subequations}
\begin{eqnarray}
	\accentset{(n)}{\mathscr{C}}_\mathrm{mass(N)} &=&
	-\frac{1}{2} \overline{N_\mathrm{R}^i} \,
	 \delta \accentset{(n)}{M}_{ij} \,
	 N_\mathrm{R}^{\ChargeC j} + \text{h.c.} \;,
\\
	\accentset{(n)}{\mathscr{C}}_{Y_\nu} &=&
	-\RaiseBrace{\bigl(}
	 \delta\accentset{(n)}{Y_\nu}\RaiseBrace{\bigr)}_{\!if} 
	\overline{N_\mathrm{R}^i}\,\widetilde \phi^\dagger \ell^f_\mathrm{L}
	+ \text{h.c.} \;,
\\
	\accentset{(n)}{\mathscr{C}}_\kappa &=&
	\frac{1}{4} \delta\accentset{(n)}{\kappa}_{gf} \,
	 \overline{\ell_\mathrm{L}^\ChargeC}^g_c \varepsilon^{cd} \phi_d\,\,
	 \ell_{\mathrm{L}b}^f \varepsilon^{ba} \phi_a + \text{h.c.} \;,
\end{eqnarray}
\end{subequations}
where the sums over $i$ and $j$ run from $1$ to $n\!-\!1$.

\begin{figure}
\begin{center}
\subfigure[%
]{\(\CenterObject{\includegraphics{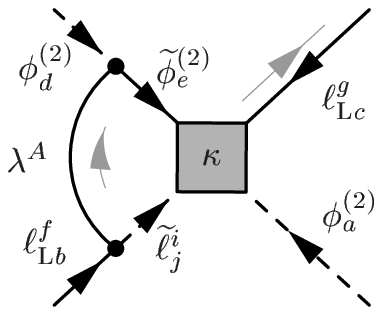}}
\)}
\hfil
\subfigure[%
]{\(\CenterObject{\includegraphics{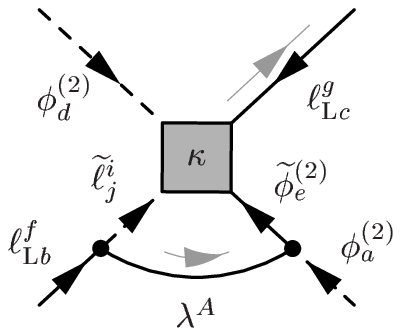}}
\)}
\hfil
\subfigure[%
]{\(\CenterObject{\includegraphics{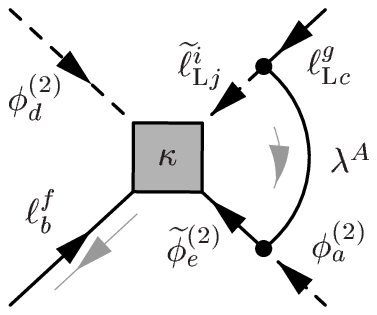}}
\)}
\\
\subfigure[%
]{\(\CenterObject{\includegraphics{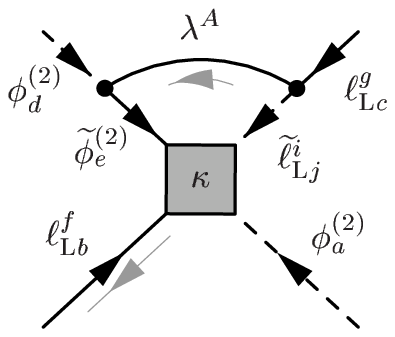}}
\)}
\hfil
\subfigure[%
]{\(\CenterObject{\includegraphics{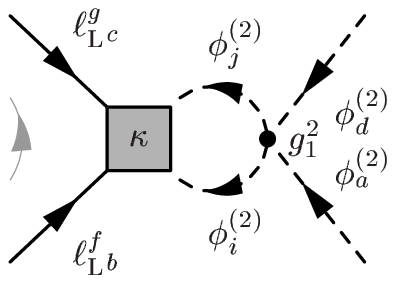}}
\)}
\hfil
\subfigure[%
]{\(\CenterObject{\includegraphics{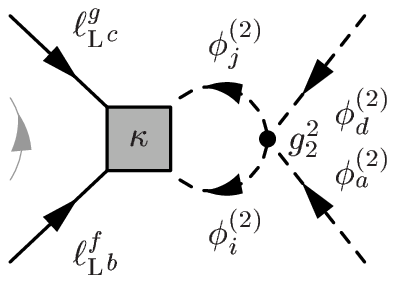}}
\)}
\end{center}
\caption{Figures (a)--(d) are the contributions from the gauginos 
 $\lambda^A$ to the renormalization of the dimension 5 operator in the
 MSSM.  Figures (e) and (f) show the $D$-term contributions.
 The gray arrow indicates the fermion flow 
 as defined in \cite{Denner:1992vz}.
}
\label{fig:MSSMContr}
\vspace{5mm}
\end{figure}

In the extended MSSM, only wavefunction renormalization is required
except for the contributions from the gauge boson - matter interactions.
Fixing the $R_\xi$ gauges and using Wess Zumino (WZ) gauge
breaks supersymmetry explicitly, and thus the non-renormalization theorem is
not manifest. Hence, the counterterms for the vertices do not vanish in general.
We use the same notation for them as in the SM. 
The relevant diagrams for 
the renormalization of the the $\accentset{(n)}{\kappa}$-vertex
are the gauge contributions similar to those of the SM, 
the gaugino contributions (figure \ref{fig:MSSMContr}~(a)--(d)) 
and the diagrams from the $D$-terms
(figure \ref{fig:MSSMContr}~(e)--(f)).
The resulting wavefunction renormalization constants are given by
\begin{subequations}
\begin{eqnarray}
\delta\accentset{(n)}{Z}_{\ell_\mathrm{L}}
& = & -\frac{1}{16\pi^2}  \left[
2\, \accentset{(n)}{Y}^\dagger_\nu \accentset{(n)}{Y}_\nu 
+ 2\,Y_e^\dagger Y_e
+ \frac{1}{2}(\xi_B - 1) g^2_1 + \frac{3}{2}(\xi_W - 1) g^2_2 \right]
\frac{1}{\epsilon} ,
\\
\delta\accentset{(n)}{Z}_{\phi^{(2)}}
& = & 
-\frac{1}{16\pi^2}\left[
2\,\Tr \RaiseBrace{\bigl(}\accentset{(n)}{Y}^\dagger_\nu \accentset{(n)}{Y}_\nu\RaiseBrace{\bigr)}
+6\,\Tr (Y_u^\dagger Y_u)
\vphantom{\frac{1}{2}} \right.
\nonumber\\
&& \left.\hphantom{-\frac{1}{16\pi^2}\left[\right.}
+ \frac{1}{2}(\xi_B + 1) g^2_1 + \frac{3}{2}(\xi_W + 1) g^2_2\right]
\frac{1}{\epsilon} \;,
\\
\delta\accentset{(n)}{Z}_{N}
& = & 
-\frac{1}{16\pi^2}\left[
4\,\accentset{(n)}{Y}_\nu \accentset{(n)}{Y}^\dagger_\nu 
\right]\frac{1}{\epsilon} \;,
\end{eqnarray}
\end{subequations}
and the vertex renormalization constants are
\begin{subequations}
\begin{eqnarray}
\delta\accentset{(n)}{Y_\nu} 
&=&-\frac{1}{16\pi^2}\left[ \frac{1}{2} (\xi_B +2) g^2_1\,\accentset{(n)}{Y_\nu}+
\frac{3}{2} (\xi_W + 2) g^2_2\,\accentset{(n)}{Y_\nu}
\right]\frac{1}{\epsilon}\;,
\\
\delta\accentset{(n)}{\kappa}
&=&
-\frac{1}{16\pi^2}\left[(\xi_B+2) g_1^2 \: \accentset{(n)}{\kappa} +
 3(\xi_W+2) g_2^2 \: \accentset{(n)}{\kappa} \,
\right]\frac{1}{\epsilon}\;,
\\
\delta\accentset{(n)}{M} &=& 0	
\;.
\end{eqnarray} 
\end{subequations}

\section{Beta-Functions in the Effective Theories}
\label{sec:beta-Ftns}

\subsection{Standard Model with Additional Majorana Neutrinos}
Using the counterterms calculated in the previous section, we find in the SM
the following $\beta$-functions $\accentset{(n)}{\beta}_\kappa = \mu
\frac{\D}{\D \mu}\accentset{(n)}{\kappa}_{gf}$ for the effective vertex
below the $n$th threshold:
\begin{eqnarray}\label{eq:RGEForKappaBetweenThresholds}
16\pi^2\accentset{(n)}{\beta}_\kappa & = & 
 -\frac{3}{2} (Y_e^\dagger Y_e)^T \:\accentset{(n)}{\kappa}
 -\frac{3}{2}\,\accentset{(n)}{\kappa} \, (Y_e^\dagger Y_e)
 + \frac{1}{2} \RaiseBrace{\bigl(} \accentset{(n)}{Y}^\dagger_\nu   
   \accentset{(n)}{Y}_\nu \RaiseBrace{\bigr)}^T \,
  \accentset{(n)}{\kappa}
 +\frac{1}{2}\,\accentset{(n)}{\kappa} \: \RaiseBrace{\bigl(}
 \accentset{(n)}{Y}^\dagger_\nu\accentset{(n)}{Y}_\nu\RaiseBrace{\bigr)}
\nonumber \\*
&& {}
 +2\,\Tr(Y_e^\dagger Y_e)\,\accentset{(n)}{\kappa} 
 +2\, \Tr \RaiseBrace{\bigl(} \accentset{(n)}{Y}^{\dagger}_\nu 
 \accentset{(n)}{Y}_\nu\RaiseBrace{\bigr)}\,\accentset{(n)}{\kappa} 
 +6\,\Tr(Y_u^\dagger Y_u)\,\accentset{(n)}{\kappa} 
  \nonumber \\*
 && {} 
 +6\,\Tr(Y_d^\dagger Y_d)\,\accentset{(n)}{\kappa}
- 3 g_2^2\: \accentset{(n)}{\kappa}
 +\lambda\accentset{(n)}{\kappa}
 \;.
\end{eqnarray}
The method used to calculate $\beta$-functions from counterterms
in MS-like renormalization schemes for tensorial quantities is described
in \cite{Antusch:2001ck}.
For the Yukawa matrix, the $\beta$-function 
$\accentset{(n)}{\beta}_{Y_\nu}$ $(n>1)$ is given by
\begin{eqnarray}\label{eq:RGEForYNuBetweenThresholds}
16\pi^2 \accentset{(n)}{\beta}_{Y_\nu}
 &=&
 \accentset{(n)}{Y}_\nu \left[ 
 	\frac{3}{2} \RaiseBrace{\bigl(}
	\accentset{(n)}{Y}^\dagger_\nu\accentset{(n)}{Y}_\nu\RaiseBrace{\bigr)}
	- \frac{3}{2}(Y_e^\dagger Y_e)
+ \Tr \RaiseBrace{\bigl(}\accentset{(n)}{Y}^{\dagger}_\nu  
\accentset{(n)}{Y}_\nu\RaiseBrace{\bigr)} +\Tr (Y_e^\dagger Y_e) \right.
\nonumber \\
&&\hphantom{\accentset{(n)}{Y}_\nu \left[ \right.} \left. 
{}+ 3\,\Tr(Y_u^\dagger Y_u)+3\,\Tr(Y_d^\dagger Y_d)
-\frac{3}{4} g_1^2 -\frac{9}{4} g_2^2 \right]
 \;. 
\end{eqnarray}
Calculating the $\beta$-function
for the Majorana mass matrix of the singlets yields
\begin{eqnarray}\label{eq:RGEForMBetweenThresholds}
16\pi^2 \accentset{(n)}{\beta}_{M} &=& \RaiseBrace{\bigl(}\accentset{(n)}{Y}_\nu   
   \accentset{(n)}{Y}^\dagger_\nu \RaiseBrace{\bigr)}\, \accentset{(n)}{M} 
   + \accentset{(n)}{M}\,\RaiseBrace{\bigl(}\accentset{(n)}{Y}_\nu   
   \accentset{(n)}{Y}^\dagger_\nu \RaiseBrace{\bigr)}^T \;.
\end{eqnarray}

\subsection{MSSM with Additional Singlets}
In the MSSM with  additional chiral superfields including 
sterile neutrinos, the $\beta$-function
for the effective vertex below the $n$th threshold is given by
\begin{eqnarray}\label{eq:RGEForKappaBetweenThresholdsMSSM}
16\pi^2 \accentset{(n)}{\beta}_\kappa & = & 
 (Y_e^\dagger Y_e)^T \: \accentset{(n)}{\kappa}
 + \accentset{(n)}{\kappa} \, (Y_e^\dagger Y_e)
 + \RaiseBrace{\bigl(} \accentset{(n)}{Y}^\dagger_\nu   
   \accentset{(n)}{Y}_\nu \RaiseBrace{\bigr)}^T\,\accentset{(n)}{\kappa}
 + \accentset{(n)}{\kappa} \: \RaiseBrace{\bigl(}
 \accentset{(n)}{Y}^\dagger_\nu\accentset{(n)}{Y}_\nu\RaiseBrace{\bigr)}
\nonumber \\
&& {} + 2\, \Tr \RaiseBrace{\bigl(} \accentset{(n)}{Y}^{\dagger}_\nu 
 \accentset{(n)}{Y}_\nu\RaiseBrace{\bigr)}\,\accentset{(n)}{\kappa}
 +6\,\Tr( Y_u^\dagger Y_u)\,\accentset{(n)}{\kappa} 
 -2 g_1^2 \:\accentset{(n)}{\kappa}- 6 g_2^2 \:\accentset{(n)}{\kappa}
 \; .
\end{eqnarray}
For $\accentset{(n)}{\beta}_{Y_\nu}$ we obtain
\begin{eqnarray}\label{eq:RGEForYNuBetweenThresholdsMSSM}
16\pi^2 \accentset{(n)}{\beta}_{Y_\nu}
 &=&
 \accentset{(n)}{Y}_\nu
 \left[ 3\, \accentset{(n)}{Y}^\dagger_\nu
\accentset{(n)}{Y}_\nu + Y_e^\dagger Y_e
+ \Tr \RaiseBrace{\bigl(}\accentset{(n)}{Y}^{\dagger}_\nu  
\accentset{(n)}{Y}_\nu\RaiseBrace{\bigr)} +3\Tr (Y_u^\dagger Y_u) 
- g_1^2 - 3 g_2^2 \right]
\end{eqnarray}
and the $\beta$-function for the Majorana mass matrix of the singlets is
\begin{eqnarray}\label{eq:RGEForMBetweenThresholdsMSSM}
16\pi^2 \accentset{(n)}{\beta}_{M} &=& 
   2\,\RaiseBrace{\bigl(}\accentset{(n)}{Y}_\nu
   \accentset{(n)}{Y}^\dagger_\nu \RaiseBrace{\bigr)}\, \accentset{(n)}{M} 
   + 2\,\accentset{(n)}{M}\,\RaiseBrace{\bigl(}\accentset{(n)}{Y}_\nu   
   \accentset{(n)}{Y}^\dagger_\nu \RaiseBrace{\bigr)}^T \;.
\end{eqnarray}

The $\beta$-functions for the gauge couplings and for the Yukawa
couplings of the quarks and charged leptons are not listed here.
We found them to be the same as in the extended SM or
MSSM \cite{Chankowski:2001mx}, if one substitutes
$Y_\nu \rightarrow \accentset{(n)}{Y}_\nu$.

\subsection{Calculation of the Low-Energy Effective Neutrino Mass Matrix}

From the above $\beta$-functions, the low-energy effective neutrino mass
matrix can now be calculated as follows: At the GUT scale, we start with
the Yukawa matrices $Y_\nu$ and the Majorana mass matrix $M$ for the
sterile neutrinos.  Using the relevant RGE's
(\ref{eq:RGEForYNuBetweenThresholds}),
\eqref{eq:RGEForMBetweenThresholds}
or (\ref{eq:RGEForYNuBetweenThresholdsMSSM}),
\eqref{eq:RGEForMBetweenThresholdsMSSM} (with the superscripts $(n)$
omitted) together with those of the gauge and the other Yukawa
couplings, we calculate the renormalization group running of $Y_\nu$,
$M$ and the remaining parameters of the theory.

At the first mass threshold, i.e.\ the largest
eigenvalue $M_{n_G}$ of $M$, we integrate out the heaviest sterile
neutrino and perform tree-level matching according to equation 
\eqref{eq:MatchingConditionKappaN}.
Note that this procedure is only possible in
the mass eigenstate basis at the threshold, which is different from the
original one at the GUT scale, since the RG evolution produces non-zero
off-diagonal entries in $M$.  Therefore, the mass matrix has to be
diagonalized by a unitary transformation, $M \to U^T M \, U$, which
leads to the redefinitions 
$N_\mathrm{R} \to U^T N_\mathrm{R}$, 
$\SuperField{\nu}^\ChargeC \to U^T \SuperField{\nu}^\ChargeC$ and 
$Y_\nu \to U^T Y_\nu$
of the singlet neutrino fields and their Yukawa matrix.%
\footnote{One could worry that the running, which spoils the diagonal
 structure of $M$, might require a constant re-diagonalization while
 solving the RGE's, since their derivations assume a diagonal mass
 matrix.  However, this is not necessary because the RGE's are invariant
 under the transformations that diagonalize $M$.
}

Integrating out the heaviest neutrino state yields an effective theory 
valid at mass scales below $M_{n_G}$. 
The effective dimension 5 operator $\accentset{(n_G)}{\kappa}$
that gives Majorana masses to the left-handed 
SM neutrinos appears in this effective theory.
Next, $\accentset{(n_G)}{Y_\nu}\,$, $\,\accentset{(n_G)}{\kappa}$,
$\accentset{(n_G)}{M}$, $Y_e$ etc.\ are evolved down to the next
threshold, the largest eigenvalue of the remaining mass matrix
$\accentset{(n_G)}{M}$.  The RGE's that determine the running of the
dimension 5 effective operator between the thresholds are given by
equation (\ref{eq:RGEForKappaBetweenThresholds}) or
(\ref{eq:RGEForKappaBetweenThresholdsMSSM}), respectively.

Again, changing to the mass eigenstate basis, integrating out the
singlet neutrino corresponding to this 
threshold and performing tree-level matching
gives another contribution to the effective dimension 5 operator.
The quantities in this effective theory are now 
evolved down to the next threshold and so on.
This procedure finally yields the low-energy effective neutrino mass
matrix.

\subsection{Running of the Mixing Angle in an Example with Two Generations}
\label{sec:HierarchicalRHExample}

\begin{figure}[h]
 \begin{center}
  \includegraphics[height=7.7cm]{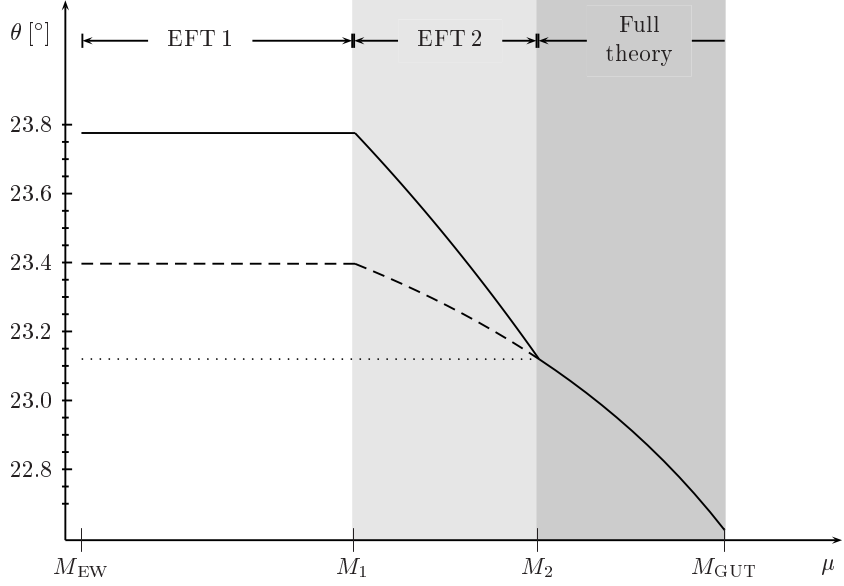}
 \end{center}
 \caption{
 RG evolution of the mixing angle $\theta$ in the extended SM with 2
 generations of lepton doublets and 2 singlets.  
 We used $M_\mathrm{GUT} = 10^{16}$ GeV and the initial
 conditions \(M_1(M_\mathrm{GUT}) = 10^8\) GeV, 
 \(M_2(M_\mathrm{GUT}) = 10^{12}\) GeV for the Majorana masses of the
 heavy neutrinos at this scale.  Besides, we chose the initial values of
 the Yukawa coupling matrices $Y_\nu(M_\mathrm{GUT})$ to be real with
 (untuned) entries between $0.025$ and $1$.
 Further explanations are given in the text.
 }
 \label{fig:ThresholdSM1}
\end{figure}

\begin{figure}[h]
 \begin{center}
  \includegraphics[height=7.7cm]{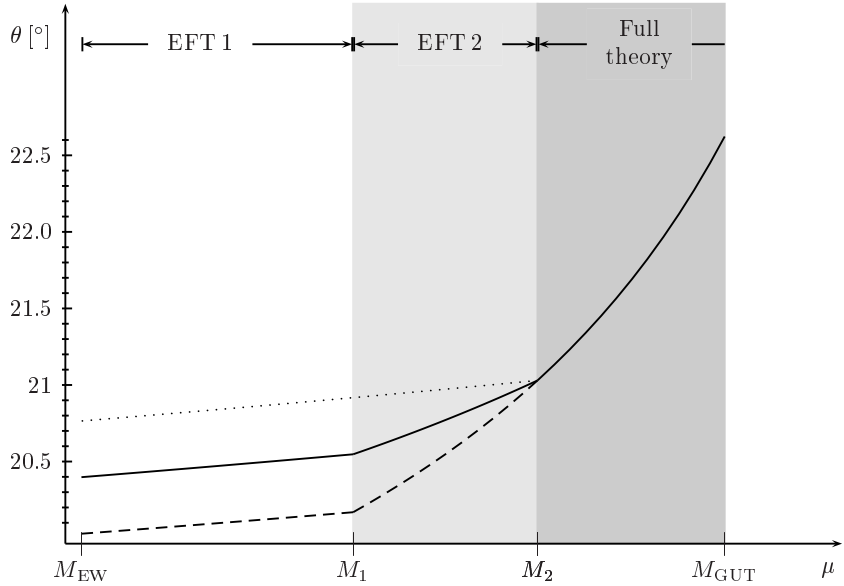}
 \end{center}
 \caption{
 RG evolution of the mixing angle $\theta$ in the extended MSSM with 2
 generations  of lepton doublets, 2 singlets and 
 $\braket{\phi^{(2)}}/\braket{\phi^{(1)}}=:\tan\beta = 35$ as well as 
 $M_\mathrm{SUSY} \approx M_\mathrm{EW}$  for simplicity. (A
 moderate change of the SUSY
 breaking scale $M_\mathrm{SUSY}$ does not change the qualitative
 picture.)  The other parameters are the same as in the SM case
 (cf.\ figure~\ref{fig:ThresholdSM1}).
 }
 \label{fig:ThresholdMSSM1}
\end{figure}

Numerical results for the RG evolution of the mixing angle $\theta$ in a
generic example with two generations of lepton doublets and two singlets
are shown as solid lines in figure
\ref{fig:ThresholdSM1} for the SM and in figure \ref{fig:ThresholdMSSM1}
for the MSSM.  Here,
\(\theta\) is defined as the angle that appears in the leptonic mixing
matrix $V = U_e^\dagger U_\nu$, where $U_e$ diagonalizes
$Y_e^\dagger Y_e$ and $U_\nu$ diagonalizes the effective
mass matrix of the active (non-sterile) neutrinos.  
Below the lowest threshold, the latter is proportional to the
coupling $\kappa$.
In the energy region where heavy neutrinos are present, the effective Majorana 
mass matrix of
the non-sterile neutrinos is given by 
$\accentset{(n)}{\kappa} + 2 \accentset{(n)}{Y}_\nu^T
\accentset{(n)}{M}^{-1} \accentset{(n)}{Y}_\nu$.

\clearpage

The transitions to the various effective theories at the mass thresholds
lead to pronounced kinks in the evolution.  For comparison, the dotted
and dashed lines in figures \ref{fig:ThresholdSM1} and
\ref{fig:ThresholdMSSM1} show the results when both heavy neutrinos are
integrated out at the higher or the lower threshold, respectively.
Obviously, this produces large deviations from the true
evolution, and the correct result need not even
lie between the two extreme cases.  Although this is only shown for the
SM in our example, the same happens in the MSSM, if suitable initial
values for the Yukawa couplings are chosen.  Consequently, the correct
running of the mixing angle cannot be reproduced by integrating out all
heavy neutrinos at some intermediate mass scale 
$M_\mathrm{int} \in [M_1, M_2]$ in general.

\section{Discussion and Conclusions}
We have calculated the RGE's for the evolution of a see-saw 
neutrino mass matrix from the GUT scale to the electroweak scale in an
extension of the SM and the MSSM by an arbitrary number of gauge
singlets with Majorana masses. These masses need not be degenerate and
can even have a large hierarchy, as pointed out in the introduction.
At each mass threshold, the corresponding sterile fermion is
integrated out, which leads to an effective intermediate theory and
affects the RG evolution of the neutrino masses, mixing angles and CP
phases.  To obtain the low-energy neutrino mass matrix from the Yukawa
and Majorana mass matrices given at the GUT scale, the RGE's for the
various effective theories have to be solved.
In a numerical analysis for two flavours and two singlets, we have found
that the renormalization group evolution of the mixing angle in the case
where the heavy degrees of
freedom are integrated out appropriately differs substantially from that
in the case where all of them are integrated out at a common scale.
The correct running can in general not even be reproduced 
by integrating out all heavy neutrinos at some intermediate mass scale. 
Obviously, similar effects exist for the RG evolution of all parameters
of a given theory, such as mass eigenvalues, mixings and CP phases.

\ack
We would like to thank M.~Drees for useful discussions. 
This work was supported in part by the 
``Sonderforschungsbereich~375 f\"ur Astro-Teilchenphysik der 
Deutschen Forschungsgemeinschaft'' and by the ``Promotionsstipendium des
Freistaats Bayern''.

\end{document}